\documentclass{elsart3p}

\usepackage{graphics}
\usepackage{graphicx,color}
\usepackage{amssymb}
\usepackage{psfrag}
\usepackage{amsmath,amsfonts,amssymb}
\usepackage{tabularx}

\DeclareMathAlphabet{\mathpzc}{OT1}{pzc}{m}{it}

\def\Ij{\mathcal{I}}
\def\auf{\bigl(}     
\def\zu{\bigr)}
\def\Auf{\Bigl(}
\def\Zu{\Bigr)}
\def\reff#1{\mbox{\rm(\ref{#1})}}
\newcommand{\intl}{\int\limits}
\def\epsp{{\epsilon^p}}
\def\eps0{\epsilon_{0}}          
\def\rateeps0{\dot{\epsilon}_{0}}
\def\sigmaT{{ \sigma_{0}}}        
\def\sigmay{{ \sigma_{y}}}
\def\dist{\tilde{d}}
\begin{document}
\begin{frontmatter}

\title{Condensation and growth of Kirkendall voids in intermetallic compounds} 

\author{Kerstin Weinberg\corauthref{label1}\thanksref{label3}} and
\ead{kerstin.weinberg@tu-berlin.de}
\author{Thomas B\"ohme\thanksref{label2}}
\ead{thomas.boehme@tu-berlin.de}

\corauth[label1]{corresponding author}
\thanks[label3]{Tel.: +49-30-314-24214; fax: +49-30-314-24499}

\thanks[label2]{Tel.: +49-30-314-26440; fax: +49-30-314-24499}

\address{Institut f\"ur Mechanik, Lehrstuhl f\"ur Kontinuumsmechanik und Materialtheorie
(LKM), Sekr. MS-2, \newline Technische Universit\"at Berlin,
Einsteinufer, 10587 Berlin, Germany}

\date{30 November 2007}

\begin{abstract}
A model for the simulation of \textsc{Kirkendall} voiding in
metallic materials is presented based on vacancy diffusion,
elastic-plastic and rate-dependent deformation of spherical voids.
Starting with a phenomenological explanation of the
\textsc{Kirkendall} effect we briefly discuss the consequences on
the reliability of microelectronics. A constitutive model for void
nucleation and growth is presented, which can be used to predict the
temporal development of voids in solder joints during thermal
cycling. We end with exemplary numerical studies and discuss the
potential of the results for the failure analysis of joining
connections.
\end{abstract}

\begin{keyword}
\textsc{Kirkendall} effect; Intermetallic compounds; Nucleation;
Vacancy diffusion; Plastic deformation

\PACS 61.72.Qq; 61.72.jd; 68.35.Fx; 66.30.-h; 62.20.F-
\end{keyword}
\end{frontmatter}

%
%
\begin{flushright}
\footnotesize{
``Get your facts first, \\
and then you can distort them.'' \\[2pt]
\it Mark Twain, (1835-1910)\rm}
\end{flushright}
\section{Introduction}\label{meso:Intro}
    
Microelectronic circuit units  consist of the functional chip unit
itself and its packaging, which includes several electro-mechanical
connections, \emph{e.g.}, solder joints between different metal
layers. Failure of these metallic components is a well established
cause of failure of the whole microelectronic system. The general
setup of a typical microelectronic system is illustrated in Figure~1
exemplarily for the Flip Chip packaging. Solder balls as well as
small, nowadays lead-free joints which are typically made of
Sn-containing alloys (\emph{e.g.}, Sn-Ag or Sn-Ag-Cu) hold the
multi-layered unit in position. In addition, the solder joints
provide electrical conductivity between the coppered layers. %
``Aging'' of the solder alloy, such as phase separation, coarsening
or the formation of InterMetallic Compounds (IMCs), as well as the
formation and growth of pores and cracks in the vicinity of
heterogeneities significantly effect the life expectation of the
joints and considerably influence the reliability of the whole
component.
    \begin{figure}[htb]
    \begin{center}
    \psfrag{F}{Si chip}
    \psfrag{V}{Voids}
    \psfrag{I1}{Cu$_3$Sn}
    \psfrag{I2}{Cu$_6$Sn$_5$}
    \psfrag{S}{solder bulk}
    \psfrag{C}{Cu pad}
    \psfrag{U}{substrate}
    \psfrag{B}{solder ball}
    \includegraphics[width=0.45\textwidth]{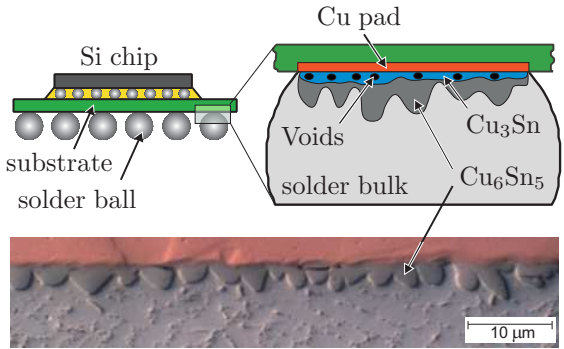}
    \caption{Intermetallic Compounds at copper-solder interfaces in microelectronic components, micrograph courtesy
            of K. M\"uller, Neue Materialien Bayreuth, Germany, 2005.\label{Pic:Intro01}}
    \end{center}
    \end{figure}

\begin{figure} [htb]
        \begin{center}
        \psfrag{V}{Kirkendall voids}
        \psfrag{K}{\textcolor{white}{Cu}}
        \psfrag{L1}{thin layer}
        \psfrag{L2}{of Cu$_3$Sn}
        \psfrag{I}{Cu$_6$Sn$_5$}
        \psfrag{1}{{\textcolor{white}{\hspace{2mm}5$\mu$m}}}
        \psfrag{C}{crack} \psfrag{2}{Sn-Ag-Cu}
        \psfrag{3}{Cu$_3$Sn-layer}
        \psfrag{4}{\textcolor{white}{Cu}}
        \psfrag{5}{\textcolor{white}{Cu$_6$Sn$_5$}}
        \includegraphics[width=7cm]{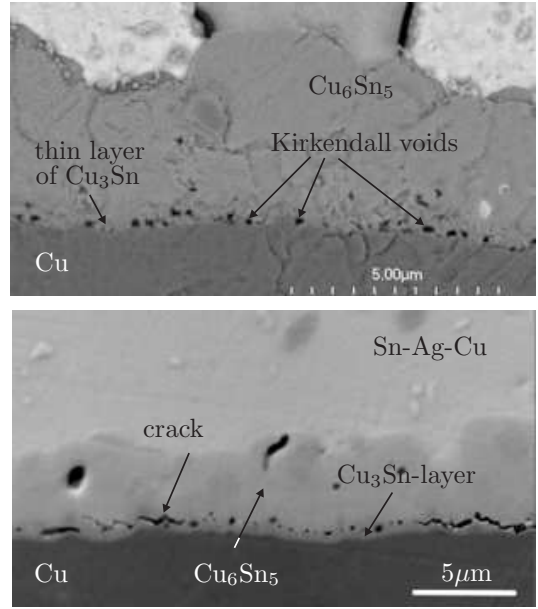}
        \end{center}
        \caption{First row: Kirkendall voids within the Cu$_3$Sn layer, photograph from \cite{Meietal2005}.
        Second row: Crack initiation by void coalescence after 1000 thermal cycles between -40$^\circ$C and
        125$^\circ$C, photograph from \cite{XuPang2005}.}
        \label{Pic:micrographs01}
    \end{figure}

During manufacturing the (molten) Sn-rich solder wets the copper pad
and IMCs are formed due to an interfacial reaction \cite{KimTu1996}.
In particular, the copper rich Cu$_3$Sn is expected to grow adjacent
to the copper substrate and Cu$_6$Sn$_5$ will form adjacent to the
Sn-based solder, \emph{cf.}, Figure \ref{Pic:micrographs01}.
However, in practise both IMCs form together with further
intermetallics to irregular shaped layers of initially 2-5 $\mu$m
hight. Due to the reflow process and to the thermal cycling during service
the solder joint ages and the IMCs grow and may reach a thickness of
20 $\mu$m and more, \emph{cf.}, \cite{XuPang2005,Prattetal1996}.
Another consequence of solder joint aging is the condensation and
growth of so-called Kirkendall voids (mainly) within the
intermetallic zones. The physics behind this mechanism may be
sketched as follows: neighboring phases or compounds change in a way
that the volume of one region grows and the volume of the another
phase reduces. In case of Sn-based solders such regions are
typically the intermetallic {compounds} Cu$_3$Sn and
Cu$_6$Sn$_5$. These IMCs show different diffusion coefficients
w.r.t. Cu and, therefore, the diffusion of Cu from the pad via the
interface Cu/Cu$_3$Sn into Cu$_3$Sn is much \emph{slower} than the
diffusion of Cu from Cu$_3$Sn into the Cu$_6$Sn$_5$ scallops,
which also cannot be ``corrected'' by the invers diffusion of Sn
through the Cu$_6$Sn$_5$/Cu$_3$ interface. Because of the unbalanced
Cu-Sn diffusion vacancy-sized voids are left which coalesce to
Kirkendall voids, \emph{cf.}, Figure \ref{Pic:micrographs01}.
Additional vacancies and defects in the crystal lattices are
generated by plastic deformation of the solder material and assist
in the process of void growth and material degradation.

The scope of this paper is to model the condensation and growth of
such voids in IMCs within a continuum mechanical framework. To this
end we idealize the material as a homogenous medium with arbitrarily
distributed vacancies and study the formation of voids and their
growth up to significant size. In the following section we outline
the constitutive equations which enables us to model the void
condensation and growth process. Subsequently, we study short term
and long term effects and present exemplary results. Finally the
temporal development of an ensemble of differently sized voids is
investigated by means of a void size distribution function.

\section{Void nucleation by vacancy diffusion}\label{section:Diffusion}

\begin{figure}[htb]
            \begin{center}
            \psfrag{L}{\textcolor{red}{vacancy diffusion}}
            \psfrag{Z}{\textcolor{black}{deformation}}
            \psfrag{M}{\textcolor{black}{void growth}}
            \psfrag{N1}{\textcolor{black}{area of attraction}}
            \psfrag{N2}{\textcolor{black}{of a central void}}
            \psfrag{c}{\footnotesize $a_0$}
            \psfrag{d}{\footnotesize $r$}
            \psfrag{s}{\footnotesize $b_0$}
            \psfrag{E}{\footnotesize $b$}
            \psfrag{F}{\footnotesize $a$}
            \includegraphics[width=0.45\textwidth]{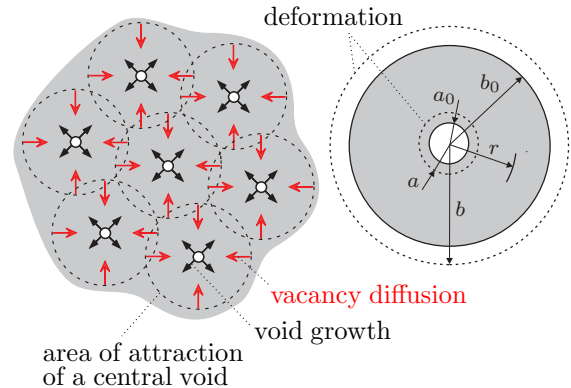}\\
        \end{center}
    \caption{Model of vacancy diffusion and model
    of a single void before and after deformation.}%
    \label{Fig:voidmodel}
    \end{figure}
To explore the feasibility of void condensation out of the scale of
crystal lattice defects as a void-nucleating mechanism we employ
here a model of \emph{vacancy condensation}. To this end let us
consider a small void possibly just a few atomic spacings in
diameter and surrounded by a supersaturated background vacancy
concentration $c_{b}$ generated, \emph{e.g.}, by unbalanced
diffusion in
the sense of Kirkendall. 
To render the problem analytically tractable, we idealize the defect
to be  spherical with characteristic radius $a$, the diffusion
constants to be isotropic, and we assume that a steady state vacancy
concentration profile is maintained all times, \emph{i.e.},
$\partial c_\mathrm{vac}/\partial t \ll
\partial J_\mathrm{vac} / \partial x$, where $J_\mathrm{vac}$ is the vacancy flux. This
eliminates the time dependence of the solution and confers spherical
symmetry to the problem. With $b$ being the (large) radius of the
basin of attraction around the void, \emph{cf.},
Figure~\ref{Fig:voidmodel}, and $b/a \to \infty$, the  diffusion
equation for \emph{volume concentration} $c$ reduces to the
following boundary value problem in spherical coordinates:
    \begin{equation}
        \frac{\partial}{\partial r}\Auf  r^2 \frac{\partial c}{\partial r}\Zu =0 \; ,
        \label{diffusionDGL}
    \end{equation}
subject to the boundary conditions
    \begin{alignat}{1}\label{diffusionDGLRB1}
    c(r={b}) &= c_{b} \; ,\\\label{diffusionDGLRB2}
    c(r=a) &= c_a = c_0 e^{d/a} .
    \end{alignat}
Here the equilibrium vacancy concentration near a free surface 
is given by
    \begin{equation}\label{meso:diffusion:c0}
    c_0 = e^{-E_v/kT}
    \end{equation}
and the concentration $c_a$ at the void surface follows as above
with $d = \frac{2 \gamma V_V}{kT}$ due to the \textsc{Gibbs-Thomson}
effect. In these expressions $\gamma$ is the surface tension, $V_V$
the atomic volume, $k$ the Bolzmann constant and $E_v$ is the
free-energy gain/loss resulting from adding a vacancy into the
system. The solution of (\ref{diffusionDGL})-(\ref{diffusionDGLRB2})
is elementary, namely
    \begin{equation}\label{meso:diffusion:Lsg}
    c(r) = c_{b} - ( c_{b} - c_a) \frac{a}{r} \;.
    \end{equation}
For void growth to take place there must be a net flux of vacancies
\emph{into} the void, which requires $c_{b} > c_a$. This in turn
requires
    \begin{equation}\label{meso:diffusion:cinfty}
        c_{b} > c_0 \exp\Auf\frac{2 \gamma V_V}{a_0 kT} \Zu.
    \end{equation}
For very small values of $a_0$ or $c_{b}$ this inequality is not
satisfied and voids fail to grow. However, the value of radius $a_0$
which equals relation \reff{meso:diffusion:cinfty} at a given value
of $c_{b}$ marks the inception of void growth, \emph{i.e.}, a
\emph{critical nucleation size}. Furthermore, the flux $J$ is
defined to be the change of volume per unit area and time,
\emph{viz.}
    \begin{equation}\label{Eq:diffflux}
    J(r)= -\frac{1}{4 \pi a^2} \frac{\mathrm{d}}{\mathrm{d}t} \left(
    \frac{4}{3} \pi a^3 \right)= - \dot{a}(r) \; .
    \end{equation}
Applying additionally \textsc{Fick}'s law $J(r)=-D_V \frac{\partial
c}{\partial r}$ to Eq. (\ref{Eq:diffflux}) yields for voids of
radius $a$ by means of Eq. (\ref{meso:diffusion:Lsg}):
    \begin{equation}\label{meso:diffusion:dota}
    \dot{a}(a) = \frac{D_V}{a} \auf c_{b} - c_0 e^{d/a} \zu \; ,
    \end{equation}
where $D_V$ is the vacancy diffusion coefficient. As the void grows,
$c_a$ decreases according to Eq.~\reff{diffusionDGLRB2} which
amplifies the concentration gradient towards the void and, thus,
accelerates the void growth. The rate of void growth may be computed
by means of Eq. (\ref{meso:diffusion:dota}).

Following the strategy of Wagner \cite{Wagner1961} but allowing for an additional
source term $s(t)$ the background vacancy concentration $c_{b}$ can be subjected to a ``void volume
balance'' of the form:
    \begin{alignat}{1}\label{diffusion:massbalance}
    &\dot c (t)  =  s(t) -  \intl_{a_\text{vac}}^{a_\text{max}}  \dist(a,
    t) \, \dot{a}(a)  \, 4 \pi a^2  \; \mathrm{d}a
    \nonumber \\
    &=
    s(t) -   \intl_{a_\text{vac}}^{a_\text{max}}   \dist(a,
    t) \, D_V \auf c_{b} - c_0 e^{d/a} \zu  4 \pi a  \; \mathrm{d}a \,
    .
    \end{alignat}
Here $\dist(a,t)$ denotes a \emph{void size distribution function} describing
the fraction  of voids with a specific size $a \in[a_\text{vac},
a_\text{max}]$ at time $t$. At this point we neglect the statistics
and consider only one void size, $\dist(a,t)=1$, but we will study different void
 distributions in Section~\ref{section:Distribution}.
The
source term $s(t)$ in Eq.~\reff{diffusion:massbalance} represents a
\emph{vacancy production rate} due to unbalanced diffusion 
which is caused by the different diffusion coefficients of the
migrating substances. Please note that
Eq.~\reff{meso:diffusion:dota} has been derived   in
\cite{FischerSvoboda2007} in a thermodynamic consistent manner (for
a constant number of vacancies). In particular, the vacancy
diffusion coefficient $D_V$ has been identified as the tracer
diffusion coefficient of vacancies.

In addition we 
ask for an energetic formulation of the
above vacancy diffusion problem. In particular, we look for a
\emph{diffusion rate potential} $\Phi (\dot a, a)$, for which the
variational form:
    \begin{equation}\label{diffusion:deltaPhi}
    \frac{\delta \Phi (\dot a, a)}{\delta \dot a} = 0
    \end{equation}
holds. For that reason we follow the strategy in
\cite{CuitinoOrtiz1995} and multiply Eq. (\ref{meso:diffusion:dota})
with a characteristic factor $E_v/D_V$. A subsequent integration
w.r.t. $\dot a$ finally results in
    \begin{equation}\label{meso:diffusion:psi}
    \Phi (\dot a, a) %
    = \frac{E_v \dot{a}^2}{2D_V} %
    - \frac{E_v \dot{a}}{a} \auf c_{b} - c_0 e^{d/a} \zu .
    \end{equation}

\section{Constitutive model of void growth}\label{section:Constitutive}
    

Once voids are nucleated diffusion is not the only mechanism which
triggers their growth within the IMCs. To set up a general
\emph{variational model} for void growth in a deforming material we
postulate the existence of a {free energy density function}
associated with the deformation of expanding voids and embedding
material. Additionally we require the time-dependent constitutive
equations to derive from power potentials. %
Thence we assume the power of the external forces acting on the
material to be \emph{completely} compensated by the change of its
free energy and its rate potentials.

Let us now ask, which energy contributions result from the
deformation of \emph{one} void subjected to \textbf{(a)}, the power
$P$ of a remote pressure $p(t)$. These are: \textbf{(b)} the energy
of the free void surface ${S}$,  \textbf{(c)} the deformation energy
${W}$ of the embedding material and  \textbf{(d)} the rate power of
creep deformation $\Psi$ and diffusion $\Phi$. Additional energy and
power contributions may play a role in specific regimes,
\emph{e.g.}, the kinetic energy in case of a very rapid loading,
\emph{cf.}, \cite{WeinbergBohme2006II}. Now an action integral
$\Ij(\dot{a})$ can be formulated as sum of all rate of energy and
power contributions. \textsc{Hamilton}'s principle simply requires
stationarity of the action integral $ \delta \Ij({\dot a}) =0 $, or,
equivalently,
    \begin{equation}\label{eq:ansatz}
    \frac{\delta} {\delta \dot{a}}  \Bigl [ \frac{\mathrm{d}}{\mathrm{d}t} \left( {W}+  {S}    \right) + \Psi + \Phi - {P} \Bigr ]=0 .
    \end{equation}

In what follows we imagine the material to be a conglomerate of
(initially very small) spherical voids each at every instance
completely embedded in the material, \emph{i.e.},  we exclude here
the process of coalescence, \emph{cf.}, Figure \ref{Fig:voidmodel}.
Consider now a void surrounded by a sphere of influence of material
with radius $b$ and let it expand for some reason. Presuming a
\emph{volume preserving deformation} it holds for all $r \in [a,b]$
    \begin{equation}\label{kirk:incompressible}
    \frac{\mathrm{d}}{\mathrm{d} t} \frac{4 \pi}{3} (r^3 - a^3 ) = 0 \;
    \Rightarrow \;
    b = \left( b_0^3 - a_0^3 + a^3 \right) ^{\frac{1}{3}},
    \end{equation}
in which the index $0$ refers to the initial state, \emph{cf.},
Figure \ref{Fig:voidmodel}. The kinematic relation
\reff{kirk:incompressible} will be employed subsequently to express
functions of radius $b$ as functions of current void radius $a$ and
the initial geometry. Furthermore, the rate of straining of the void
surrounding material can be defined as
    \begin{equation}\label{epspdot}
    \dot \epsilon \stackrel{(\mathrm{def})}{=} \frac{\partial \dot r}{\partial r}  =
    \frac{\partial }{\partial r} \Auf \frac{a^2}{r^2} \dot{a}\Zu = \frac{2 { a^2}}{r^3}\dot {{ a}}
    \; ,
    \end{equation}
where  $r$ is the radius of  void surrounding material, see
Figure~\ref{Fig:voidmodel}.

In what follows we consider the different energy contributions
\textbf{(a-d)} in detail.

\subsection{External power}

The external power put into the system by an applied (positive or
negative) pressure $p(t)$ reads
    \begin{equation}
        P_\mathrm{tot}= \frac{\mathrm{d}}{\mathrm{d}t}\int_V p(t) \,d
        V \nonumber
    \end{equation}
and for one void we obtain
    \begin{equation}\label{dotWexSphere}
        {{P}}(\dot a, a) =  p(t) \,  4 \pi a^2  \dot a.
    \end{equation}
In the solder material under consideration here the effect of
thermal cycling on the growth of voids is of particular interest. To
this end temperature cycles $T(t)$ are considered to strain the
material and to induce a pressure of the form
    \begin{equation}
        p(t) = \kappa^*(a) \; 3 \alpha \; [T_0-T(t)]
    \end{equation}
where $\kappa^*$   is the effective bulk modulus and $\alpha$ is the
thermal expansion   coefficient.  Following a classical approach of
homogenization, \textit{cf.}, \cite{LeeWesterman1970,Hashin1983}, we
obtain for an assemblage of spherical voids enclosed in an isotropic
material with bulk modulus $\kappa$ and shear modulus $\mu$   the
effective bulk modulus $\kappa^*$ as
    \begin{equation}\label{Hashin:Kstern}
        \kappa^* = \kappa \left( 1 - \frac{a^3}{b^3} \frac{3\kappa + 4 \mu}{3 \kappa + 4\mu b^3/a^3} \right) .
    \end{equation}

\begin{figure}[htb]
    \begin{center}
    \psfrag{K}{\footnotesize $T$ in $^\circ$C}
    \psfrag{T}{\footnotesize time $t$ in min}
    \includegraphics[width=0.45\textwidth]{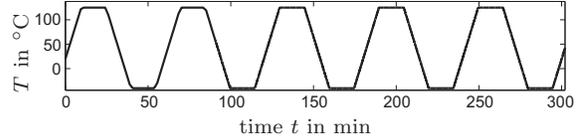}
    \caption{Thermal cycling between -40$^\circ$C and 125$^\circ$C, cycles of 60 min with 15 min hold time.}\label{temperatureCycles}
    \end{center}
    \end{figure}

The temperature cycles between $-40^\circ$C and $125^\circ$C within
one hour, in which 15 minutes at up soak and low soak temperature,
respectively, see Figure~\ref{temperatureCycles}.
    %

\subsection{Surface energy}

The surface energy of one void with radius $a(t)$ is written as
    \begin{equation}\label{SSphere}
        {S}(a) = 4 \pi a^2 \gamma \; ,
    \end{equation}
where $\gamma$ is the surface energy per unit undeformed area [N/m].

\subsection{Elastic-plastic deformation energy}

The deformation energy for the elastic-plastic material response can
be derived as follows. We presume a \textsc{Ramberg-Osgood} power
law,
    \begin{equation}\label{sigmay}
    \epsilon = \epsilon^e +
    \frac{\sigmaT}{E}\Auf\frac{\sigma}{\sigmaT}\Zu^{n} ,
    \end{equation}
where $\sigma$ and $\epsilon$ are the effective stress and strain,
respectively. Furthermore $\epsilon^e $ denotes the elastic strain
component and $E \equiv E(T)$ and $\sigmaT \equiv \sigmaT(T)$
represent temperature dependent \textsc{Young}'s modulus and initial
yield stress. Exponent $n \in [1,\infty)$ determines the
stress-strain curve; in particular $n=1$ prescribes linear
elasticity, whereas $n \to \infty$ enforces perfect (rigid)
plasticity. In order to resolve Eq.~\reff{sigmay} let us assume the
elastic strain to be given and decompose $\epsilon = \epsilon^e +
\epsilon^p$. Then the dissipated energy of the deformation per unit
volume can be computed from
    \begin{alignat}{1}\label{wp}
    \int_0^{t} \sigmay \dot \epsp \; \mathrm{d} \,\bar{t} &= \int_0^{\epsp}
    \sigmay \; \mathrm{d} \bar{\epsp} \nonumber \\
    &= \frac{n \sigmaT}{n+1}  \left( \frac{E}{\sigmaT}\right)^{\frac{1}{n}}
      \auf  {\epsilon - \epsilon^e} \zu^\frac{n+1}{n}
    \end{alignat}
and with reference strain $\eps0=\sigmaT/E$ the dissipated deformation energy for one
void with surrounding material follows as
    \begin{equation}
    {W}(a; T) =  \int_a^{b}   \frac{n \sigmaT \eps0}{n+1} \left(
    \frac{\epsilon - \epsilon^e}{\eps0} \right)^\frac{n+1}{n} 4\pi r^2
    \mathrm{d}r \; . \label{WpvolSphere}
    \end{equation}
Let us now refer to the kinematic relation in Eq.~\reff{epspdot} and
assume that the void radius history ${a}(t)$  grows monotonically
from $a_0$ to $a_1$, then decreases monotonically from $a_1$ to
$a_2$, and so on. The integration of Eq. \reff{epspdot} with respect
to the time gives
    \begin{alignat}{1}
    \nonumber \epsilon(r_0,t) &= \frac{2}{3} \log \left( \frac{{ a_1}^3 +
    r_0^3
    -a_0^3}{r_0^3} \right) \\&+ %
    \frac{2}{3} \log \left( \frac{{ a_1}^3 + r_0^3 -a_0^3}{{
    a_2}^3 + r_0^3 - a_0^3} \right) \nonumber \\&+ %
    \frac{2}{3} \log \left( \frac{{ a_3}^3 + r_0^3 -a_0^3}{{ a_2}^3 +
    r_0^3 - a_0^3} \right) + \cdots \,,
    \end{alignat}
and grouping terms corresponding to increasing and decreasing intervals
yields
    \begin{alignat}{1}
    \epsilon(r_0,t) &= \frac{2}{3} \log \left( \frac{{ q(t)}^3 + r_0^3
    -a_0^3}{r_0^3} \right) \nonumber \\ &- %
    \frac{2}{3} \log \left( \frac{{ a }^3(t) + r_0^3 -a_0^3}{{ q}^3 (t) +
    r_0^3 - a_0^3} \right) , \label{epspSum2}
    \end{alignat}
where $q(t)$ is the maximum radius attained by voids of the current size $a$,
    \begin{equation}\label{epspSumq}
        q(t) = \max_{0 \le \tau \le t}  a (\tau) \, ,
    \end{equation}
\emph{i.e.}, $q(t)$ is a monotonically increasing function for every
history of ${a}(t)$. Then expression in Eq.~\reff{epspSum2} can be
summarized, \emph{viz.}
    \begin{equation}\label{epspSum}
    \epsilon(r_0,t) = \frac{2}{3} \log \left[ \frac{\auf {
    q }^3(t) + r_0^3 -a_0^3 \zu^2}{r^3_0 \; \auf { a}^3 (t) + r_0^3 -
    a_0^3 \zu } \right]
    \end{equation}
and with Eq.~\reff{kirk:incompressible} the dissipated deformation energy  in a 
shell surrounding the void is
    \begin{alignat}{1}
    &{W}(a,q; T) \; = \; 4\pi \int_{a_0}^{b_0}   \frac{n \sigmaT \eps0}{n+1}
    \Bigg[ \frac{2}{3\eps0} \; \times \nonumber \\
    & \times \; \log
    \left(\frac{\left(q(t)^3+r_0^3-a_0^3\right)^2}{r_0^3\left(a(t)^3+r_0^3-a_0^3\right)}\right)
    - \frac{\epsilon^e}{\eps0} \Bigg]^\frac{n+1}{n} r_0^2
    \;\mathrm{d}r \; .
    \label{WpvolSphereq}
    \end{alignat}
It is worth mentioning that the function in Eq. (\ref{WpvolSphereq})
is trackable by analytical means only in the special case of $n \to
\infty$. Moreover, in moderate hardening materials we know the
elastic strain component to be $\epsilon^e \approx \eps0$. Therefore
we set the last term in brackets to one but guarantee the energy to
be not negative, \emph{i.e.}, $W \ge0$.
%

\subsection{Rate-dependent deformation (Creep)}

The creep-power potential is formulated in order to capture the
effect of the rate of deformation on the rate of straining.
Experimental observations reported, \textit{e.g.}, in
\cite{Freund1993}, indicate that the material near a void is
subjected to a state of stress that is likely to cause power-law
creep, \textit{i.e.},
    \begin{equation}\label{eq:sigmarate}
    \dot{\epsilon} = \Auf \frac{\sigmay}{\sigmaT} \Zu^m \dot{\epsilon_c} \exp(-Q_c/RT) \;,
    \end{equation}
where $\dot{\epsilon}$ is the strain rate, $m$ is a creep exponent
and $\dot{\epsilon_c}$ is a  material constance. The temperature
dependence of the strain rate is controlled by the thermal
activation energy $Q_c$. Here we summarize the last terms in
Eq.~\reff{eq:sigmarate} to a \emph{reference strain rate},
$\rateeps0 = \dot{\epsilon_c} \exp(-Q_c/RT)$, with small values of
$\rateeps0$ corresponding to creep dominated deformation and
$\rateeps0 \to \infty$ to a time independent behavior. The
creep-power potential per unit volume is  defined by
    \begin{multline}\label{psim}
    \int_0^{\dot{\epsilon}} \sigmay \, \mathrm{d} \,\bar{\dot{\epsilon}}
    =  \frac{m \sigmaT \dot{\epsilon}}{m+1} \left[ \left( \frac{\dot{\epsilon}}{\rateeps0} + 1 \right)^\frac{m+1}{m} - 1 \right]\, .
    \end{multline}
For simplicity we assume now a linear rate dependence, $m=1$, and by
integration over the volume follows the creep potential for one
spherical shell as
    \begin{alignat}{1}\label{PsistarSphere}
        {\Psi}(\dot a, a) &= \int_a^b %
    \frac{\sigma_0 }{ 2\dot{\eps0}}
            \left( \frac{2 a^2 |\dot{a}|}{ r^3} \right)^2  4\pi r^2
            \, \mathrm{d} r \nonumber \\
            &=
    \frac{\sigma_0 }{ \rateeps0} \int_{a_0}^{b_0} %
            \left( \frac{a^2 |\dot{a}|}{a^3 + r_0^3 -a_0^3} \right)^2  8\pi
            r_0^2 \, \mathrm{d} r_0 \nonumber \\
            &= \frac{2\sigma_0 }{ \rateeps0}\frac{4 \pi a^3}{3}
            \Big|\frac{\dot{a}}{a}\Big|^2
            \left( 1-\frac{a^3}{a^3 + b_0^3 -a_0^3} \right),
    \end{alignat}
where we again make use of Eq.~\reff{epspdot}.

Setting  the external power equal the ``sum of the internal powers''
according to  ansatz~\reff{eq:ansatz} yields an \emph{ordinary
differential equation} for the void size $a(t)$. In particular, for
growing voids and for $n \to \infty$ in Eq. \reff{WpvolSphereq} we
obtain the expression
    \begin{alignat}{1}
    0 = &- 4 \pi p(t) a^2 + \frac{8 \pi}{3}\sigmaT a^2  \log\left(\frac{ b_0^3 - a_0^3 + a^3}{a^3}\right)
    \nonumber \\
    &+ \frac{16 \sigma_0 \pi}{ 3 \rateeps0}{ a^4} \dot{a} \Auf \frac{1}{a^3}- \frac{1}{b^3} \Zu
        \nonumber \\
    &+ 8 \pi a \gamma  \; + \frac{E_v \dot{a}}{D_V} - \frac{E_v }{a} (c_{b} -
    c_a) \; ,\label{masterequation}
    \end{alignat}
which can be solved (numerically) for {all} different void sizes of
interest. We will outline selected results subsequently.

\section{Nucleation of voids out of small defects} %

At first we study the nucleation of voids, \emph{i.e.}, the
formation of pores out of vacancy sized defects. It can easily be
seen from Eq.~\reff{masterequation} that for  very small values of
$a_0$ the diffusion term dominates. In the initial stages of void
growth the elastic-plastic material response is of minor influence.
As well we neglect the external loading for a start and the equation
of motion reads:
    \begin{equation}\label{masterequation:diffusion}
    \dot{a} =  \frac{D_V}{E_v}\Bigl[ 8 \pi a \gamma  - \frac{E_v}{a} (c_{b} -
    c_0 e^{d/a}) \Bigr] \; .
    \end{equation}
Thus the void condensation process is driven by an interplay of
vacancy diffusion and surface energy contributions. In particular,
for Eq. \reff{meso:diffusion:cinfty} to be satisfied the initial
size of the defect must exceed a critical nucleation radius which is
here approximately 2-3 times the vacancy size ($0.25$~nm). To keep
voids growing a vacancy production source is required, otherwise the
voids either collapse or reach a steady state --- depending on the
magnitude of surface tension.
In the material under consideration here the vacancy production
results from the unbalanced diffusion, \emph{i.e.}, the different
diffusion coefficients of the migrating substances,
\cite{CahnHilliard1971}. In a first approximation its rate is
assumed to be proportional to the difference of the diffusion
coefficients:
    \begin{equation}\label{CkirkDD}
        s(t) = c_K(t) \auf D_{\mathrm{Cu}} - D_{\mathrm{Sn}} \zu \; ,
    \end{equation}
where $D_{\mathrm{Cu}}$ and $D_{\mathrm{Sn}}$ are the diffusion
coefficients of copper into tin and vice versa, $c_K$ is a weighting
factor with unit [m$^2$] which accounts for progressing IMC growth.
Note that plastic deformation is an additional source of vacancy
production and factor $c_K$ may also depend on time via the rate of
plastic straining.

    \begin{figure}[htb]
    \begin{center}
    \psfrag{A}{\footnotesize $a/a_0$}
    \psfrag{B}{\footnotesize time $t$ in h}
    \psfrag{C}{\footnotesize $a_0=0.5\,\mathrm{nm}$}
    \psfrag{D}{\footnotesize $a_0=500\,\mathrm{nm}$}
    \psfrag{E}{\footnotesize $a_0=5\,\mathrm{nm}$}
    \includegraphics[width=0.38\textwidth]{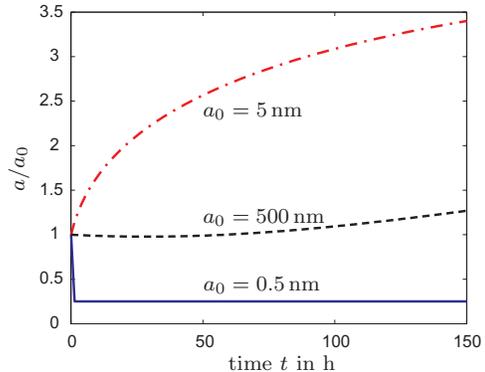}
    \caption{Condensation and growth of voids with different initial radii $a_0$
    driven by an interplay of diffusion and surface energy
    contributions.}\label{fig:diff3a0}
    \end{center}
    \end{figure}

    \begin{table}[htb]
    \renewcommand{\arraystretch}{1.3}
    \caption{Material constants for diffusion}\label{MaterialConstantsDiff}
    \centering \vspace{1ex}
    \begin{tabularx}{\linewidth}{l|X|X|X|X|X}
    \hline\hline
    $\; D_{V}$ [m$^2$/s]&  $\;c_{0}$ & $\;c_{b}$ & $\;d$ [m] & $\;s$ [1/s] & $\;\gamma$ [N/m]\\
    \hline
    $\; 10^{-17}$ & $\; 10^{-6}$ & $\;10^{-4}$ & $\;5\cdot 10^{-6}$ &$\;10^{-7}$ & $\;1$\\
    \hline\hline
    \end{tabularx}
    \end{table}

Although Cu-Sn intermetallics are of great interest for the  electronic packaging industry
the available material data on such compounds vary considerably,
see, e.g.,
 \cite{Chao_etal2007_diffCuSnIMCpdf,Galyon2005,ShackelfordB} 
and references therein. With typical 
values summarized in Table~\ref{MaterialConstantsDiff} we
obtain for the solution of Eq.~\reff{masterequation:diffusion} the results\footnote{The ODEs of Eq.
(\ref{masterequation:diffusion}) have been solved numerically  using the
Matlab$^\circledR$ solver \texttt{ ODE45 } as well as 
an explicit 2th order \textsc{Runge-Kutta} procedure.} 
displayed in
Figure~\ref{fig:diff3a0}. Small defects with an initial radius of
$a_0=0.5$~nm (twice the size of a vacancy in copper) collapse
immediately. Defects of size greater than nucleation size will grow
with a rate of void growth depending on the vacancy production rate.
On the other hand, big voids, \emph{e.g.}, $a_0=500\, \mathrm{nm}$,
basically fail to grow, here the influence of diffusion is to small
for significant void growth.

\begin{figure}[htb]
    \begin{center}
    \psfrag{A}{\footnotesize $a/a_0$}
    \psfrag{B}{\footnotesize time $t$ in h}
    \psfrag{C}{\footnotesize $a_0=500$\,nm, no load}
    \psfrag{D}{\footnotesize $a_0=5$\,nm, thermal cycling}
    \psfrag{F}{\footnotesize $a_0=5$\,nm, no load}
    \psfrag{E}{\footnotesize $a_0=500$\,nm, thermal cycling}
    \includegraphics[width=0.38\textwidth]{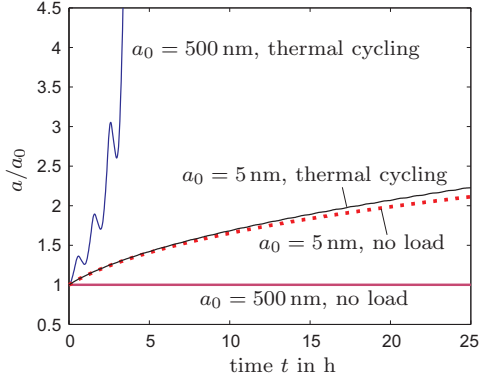}
    \caption{Influence of the external load induced by subjecting the material to
    the temperature cycles of Figure~\ref{temperatureCycles} on the diffusion driven void
    growth of two voids.}\label{fig:DiffSWx25h2a0}
    \end{center}
    \end{figure}

Figure~\ref{fig:DiffSWx25h2a0} shows the same void growth model but
now \emph{with} external loading, \emph{i.e.}, the voids are
subjected to a pressure history induced by thermal cycling. On
relatively small voids this pressure has little effect, their growth
is in first instance given by diffusional effects. However, if the
void size reaches a significant magnitude the void grows unbounded
-- a clearly non-physical effect -- which shows the necessity to
incorporate elastic-plastic deformation into the model. This effect
should be considered in the following section.
\section{Growth of voids due to thermal cycling} %

\begin{figure}[htb]
    \begin{center}
    \psfrag{A}{\footnotesize $a/a_0$}
    \psfrag{B}{\footnotesize time $t$ in min}
    \psfrag{E}{\footnotesize $a_0=50\,\mathrm{nm}$}
    \psfrag{G}{\footnotesize $n\rightarrow\infty$}
    \psfrag{H}{\footnotesize $n=5$}
    \psfrag{C}{\footnotesize $n\rightarrow\infty$}
    \psfrag{I}{\footnotesize (perfect plastic)}
    \psfrag{F}{\footnotesize $n=5$}
    \psfrag{D}{\footnotesize $a_0=50\,\mathrm{nm}$}
    \psfrag{E}{\footnotesize $a_0=500\,\mathrm{nm}$}
    \includegraphics[width=0.38\textwidth]{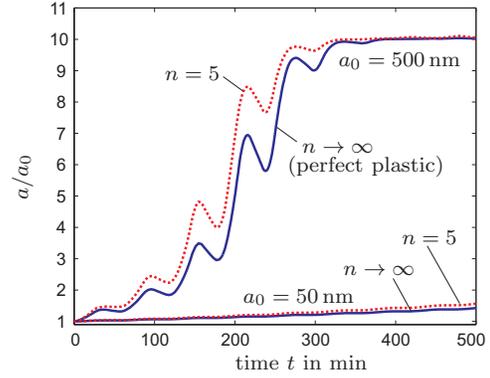}
    \caption{Growth of a medium sized and large void induced by thermal
    cycling for two elastic-plastic material laws according to Eq. \ref{sigmay}:
    $n=5$
    and perfect plasticity.}\label{fig:DiffSWexWp500min2a02n}
    \end{center}
    \end{figure}

The full model of Section~\ref{section:Constitutive} 
is now studied for medium sized and big voids with material
parameter given in Table~\ref{MaterialConstants}. Note that the
elastic-plastic deformation energy contribution in
Eq.~\reff{masterequation} is simplified in order to get an
analytical expression; a computation of the full model requires a
numerical integration of Eq. \reff{WpvolSphereq}. In
Figure~\ref{fig:DiffSWexWp500min2a02n} we see the different rates of
growth for medium sized and big voids. Here we neglect creep effects
at first, \emph{i.e.}, $\rateeps0 \to \infty$.
    \begin{table}[htb]
    \renewcommand{\arraystretch}{1.3}
    \caption{Elastic-plastic material constants\protect\footnotemark}\label{MaterialConstants}
    \centering
    \begin{tabularx}{\linewidth}{X|X|X|X|X}
    \hline\hline
    $\;E$ [GPa]&   $\;\kappa$ [GPa]& $\;\mu$ [GPa]& $\;\sigmaT$ [MPa]& $\;\alpha$ [1/K]\\
    \hline
    $\;100$&$\;80$& $\;50$& $\;450$ & $\;19\cdot 10^{-6}$  \\
    \hline\hline
    \end{tabularx}
    \end{table}
\footnotetext{Experimental data courtesy of Prof. M\"{u}ller, Inst.
f. Mech. (LKM), TU Berlin, Germany.} \addtocounter{footnote}{-1}
\stepcounter{footnote}
Both void sizes are subjected to the same thermal cycling of
Figure~\ref{temperatureCycles}, but their evolution history depends
strongly on the initial void size. In particular, smaller voids
clearly grow slower than their initially big companions. That
supports the experimental observation of several (relatively) large
\textsc{Kirkendall} voids in IMCs.
The differences in  the void size \emph{vs}. time response for the
two cases of elastic-plastic material behavior ($n=5$) and perfect
plastic approximation ($n \to \infty$) are small, the final void
size is basically determined by the external loading. Therefore, a
simplified approximation as given in Eq.~\reff{masterequation} seems
to be justified. Let us point out that the diffusion effects
invoke a ``smoothening'' of the void growth curves. Even if the underlying
elastic-plastic theory is time independent, the material
response has no sharp edges as the load history would suggest.

\begin{figure}[htb]
    \begin{center}
    \psfrag{A}{\footnotesize $a/a_0$}
    \psfrag{B}{\footnotesize time $t$ in min}
    \psfrag{C}{\footnotesize $\rateeps0=10^{-6}$}
    \psfrag{D}{\footnotesize $\rateeps0=10^{-5}$}
    \psfrag{E}{\footnotesize $\rateeps0=10^{-4}$}
    \psfrag{F}{\footnotesize $\rateeps0=10^{-3}$}
    \psfrag{G}{\footnotesize time-independent}
    \includegraphics[width=0.38\textwidth]{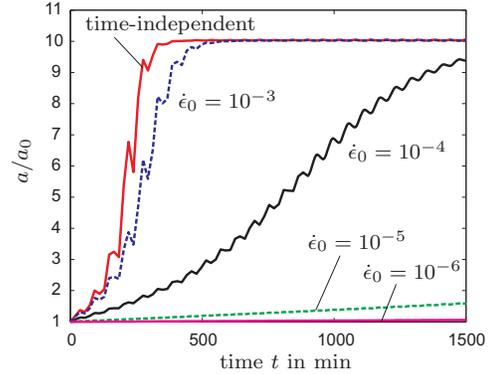}
    \caption{Effect of creep  on the growth of voids of size $a_0=500$nm for different reference strain rates.}\label{fig:PsiWpWx1500min5eref}
    \end{center}
    \end{figure}

Figure~\ref{fig:PsiWpWx1500min5eref} displays the void size
\textit{vs.} time for different reference strain rates. Small values
of  $\rateeps0$ clearly damp the evolution of voids. However, after
enough time has passed the final void size will reach the same value
as in the time-independent case as can be seen from the curves for
$\rateeps0=10^{-4}$ and $\rateeps0=10^{-3}$ in
Figure~\ref{fig:PsiWpWx1500min5eref}.

\section{Evolution of a Void Ensemble}\label{section:Distribution} %

In order to investigate an \emph{ensemble} of voids with different
initial radii we introduce -- as already indicated in Section
\ref{section:Diffusion} -- a void size distribution function
$\dist(a,x,t)$. To derive an evolution equation for the void
distribution  we make use of a mesoscopic concept described in
detail, \emph{e.g.}, in
\cite{Papenfuss2003,WeinbergBohme2006I,WeinbergBohme2006II}. In this
work we do not account for a dependence of the void distribution on
the spacial position $x$, and set
$\dist(a,x,t)=\dist(a,t)\equiv\dist$. Then we establish for a
constant number of voids a balance equation for the void size
distribution $\dist$  in the following form:
    \begin{equation}
    \frac{\partial \dist}{\partial t} +
    \frac{\partial}{\partial
    x} [ \dist \, v(x,t) ]
    + \frac{\partial}{\partial
    a} \left[ \dist \, \dot{a}(a) \right] = 0 \; .
    \label{Eq:DistrBalance}
    \end{equation}
The value of $\dot{a}(a)$ can be calculated from the constitutive
model introduced in Section \ref{section:Diffusion} and
\ref{section:Constitutive}. Moreover we neglect the spatial velocity
of the material element under consideration, $v\doteq0$, so that the
second term in Eq. (\ref{Eq:DistrBalance}) vanishes.

\begin{figure}[htb]
            \begin{center}
            \psfrag{A}{\footnotesize \hspace{-0.0cm}\vspace{5cm} $a$ in $\mu$m}
            \psfrag{B}{\footnotesize \hspace{-0.6cm}distribution $\dist$}
            \psfrag{U}{\footnotesize initial}
            \psfrag{V}{\footnotesize Eq. (\ref{Eq:DistrBalance})}
            \psfrag{W}{\footnotesize Eq. (\ref{Eq:quasiLinForm})}
            \includegraphics[width=0.29\textwidth]{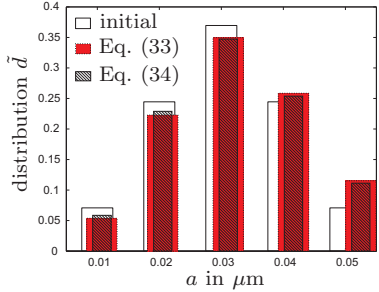}
        \end{center}
    \caption{Comparison of the numerical solution of Eq. (\ref{Eq:DistrBalance}) and (\ref{Eq:quasiLinForm})
    for 5 different radii after 2 thermal cycles.}%
    \label{Fig:VglLinFull}
    \end{figure}

To solve  Eq.~(\ref{Eq:DistrBalance}) numerically, the spatial and
temporal derivatives were discretized  by \emph{finite Differences}.
The solution of  Eq.~(\ref{masterequation}) is obtained by means of
a 2nd order \textsc{Runge-Kutta} method. For ease of computation we
also solved the quasi-linear pendant of Eq. (\ref{Eq:DistrBalance}),
\emph{viz.}
    \begin{equation}
    \frac{\partial \dist}{\partial t} +
    \dot{a}(a) \frac{\partial \dist}{\partial a} = 0 \; .
    \label{Eq:quasiLinForm}
    \end{equation}
As is shown exemplarily in Figure \ref{Fig:VglLinFull} both
equations yield almost identical results. However, to solve the full
Eq.~(\ref{Eq:DistrBalance}) a significant finer spatial discretization is
required. Therefore we proceed here with  solutions of
Eq.~(\ref{Eq:quasiLinForm}). Furthermore, for a better understanding
of the results we neglect any
void production, \emph{i.e.}, $s(t)\doteq0$ in Eq.~(\ref{diffusion:massbalance}).
However, there is no extra effort to include the
production term into the simulations and in Eqs.~(\ref{Eq:DistrBalance}) and ~(\ref{Eq:quasiLinForm}).

    \begin{figure}[htb]
                \begin{center}
                \psfrag{A}{\footnotesize \hspace{-0.0cm}\vspace{5cm} $a$ in $\mu$m}
                \psfrag{B}{\footnotesize \hspace{-0.6cm}distribution $\dist$}
                \includegraphics[width=0.34\textwidth]{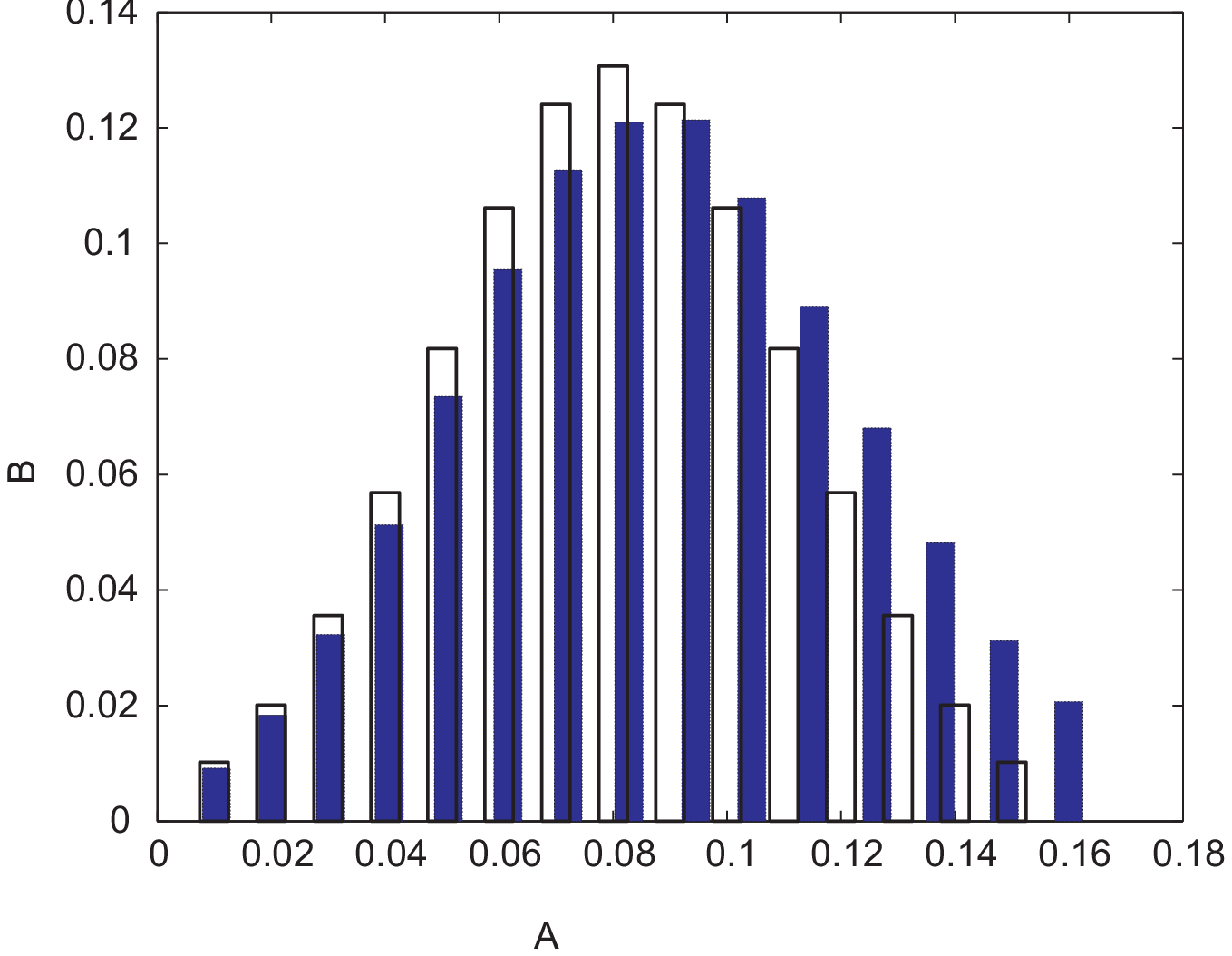}\\[5pt]
                \includegraphics[width=0.34\textwidth]{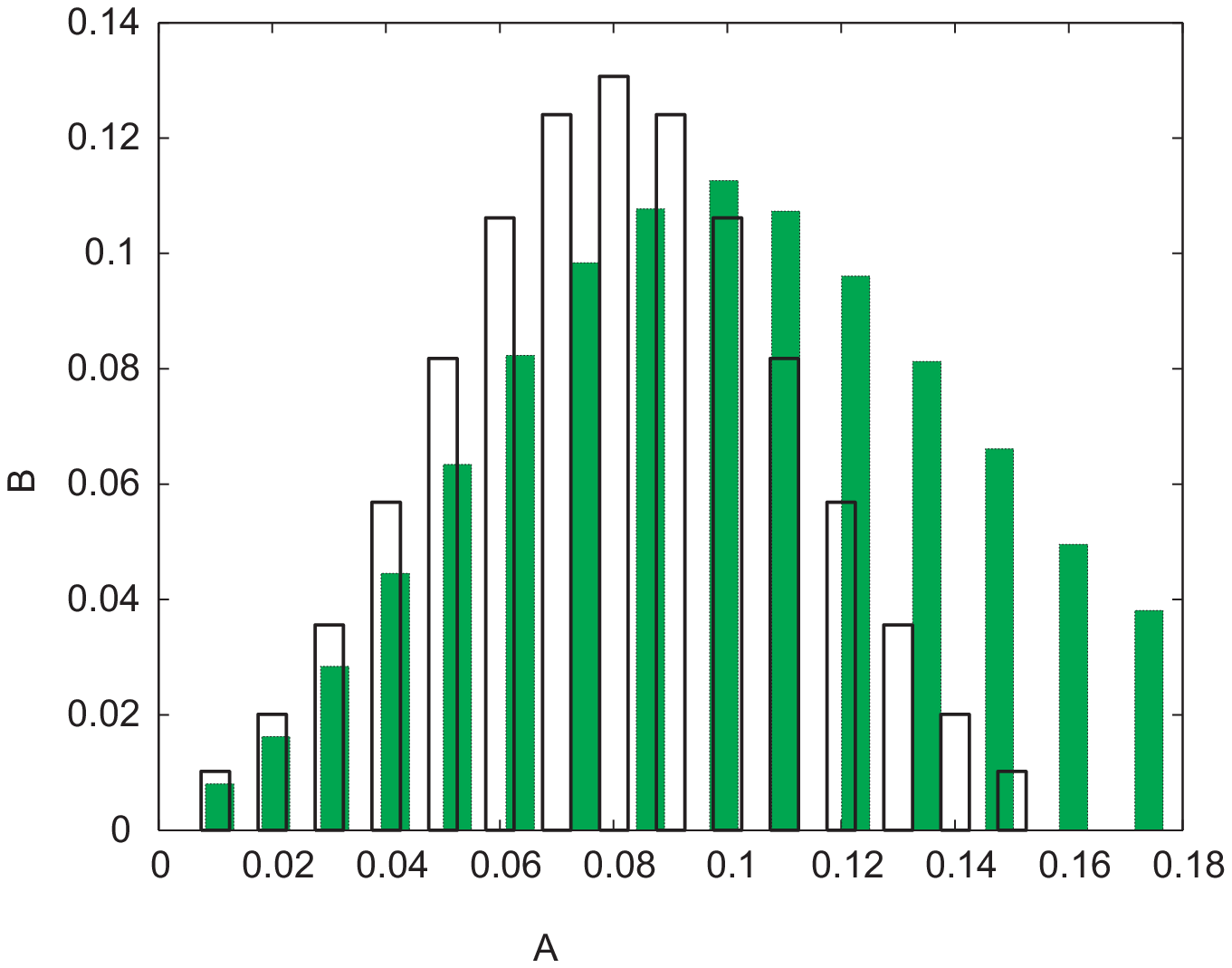}\\[5pt]
                \includegraphics[width=0.34\textwidth]{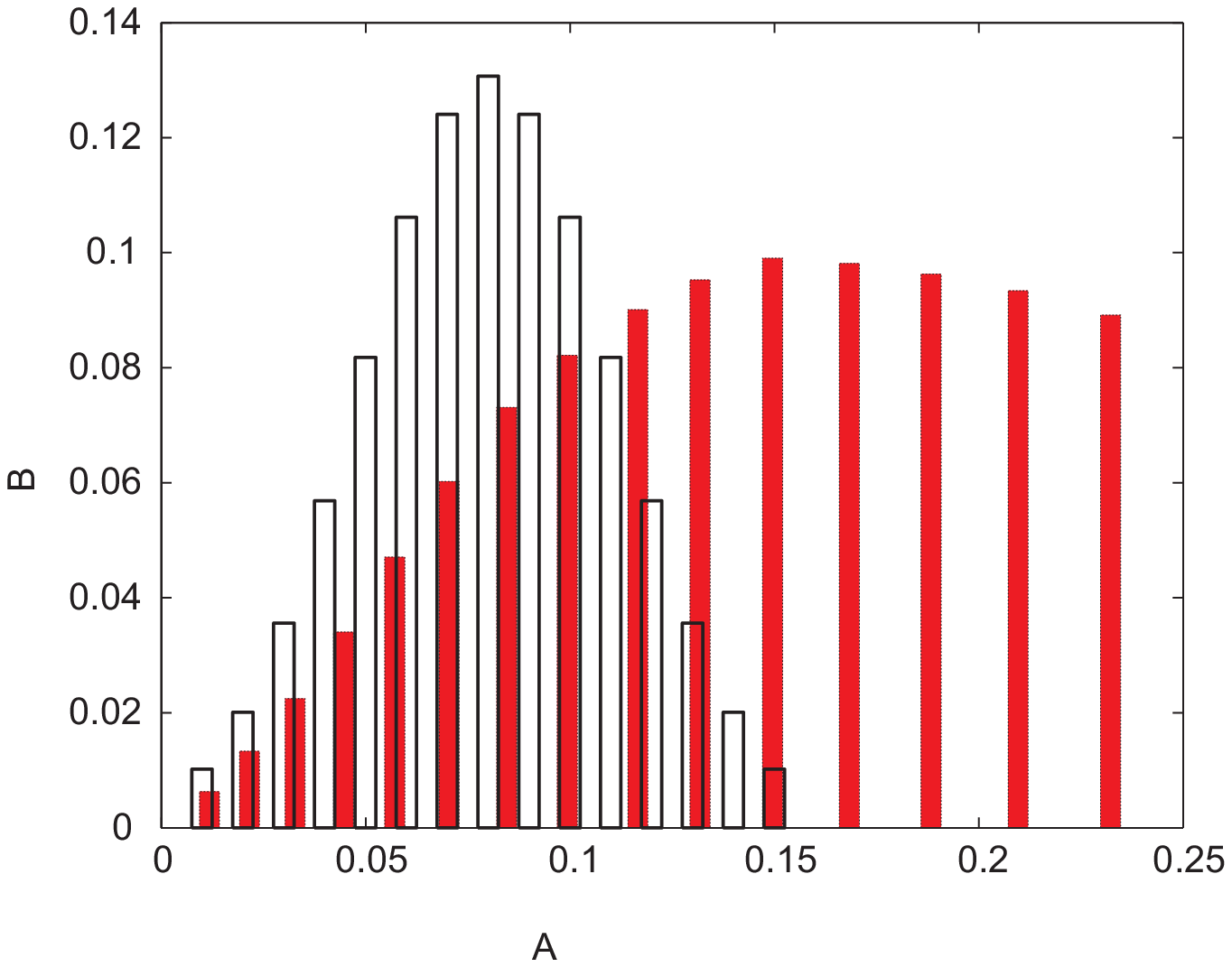}\\[5pt]
                \includegraphics[width=0.34\textwidth]{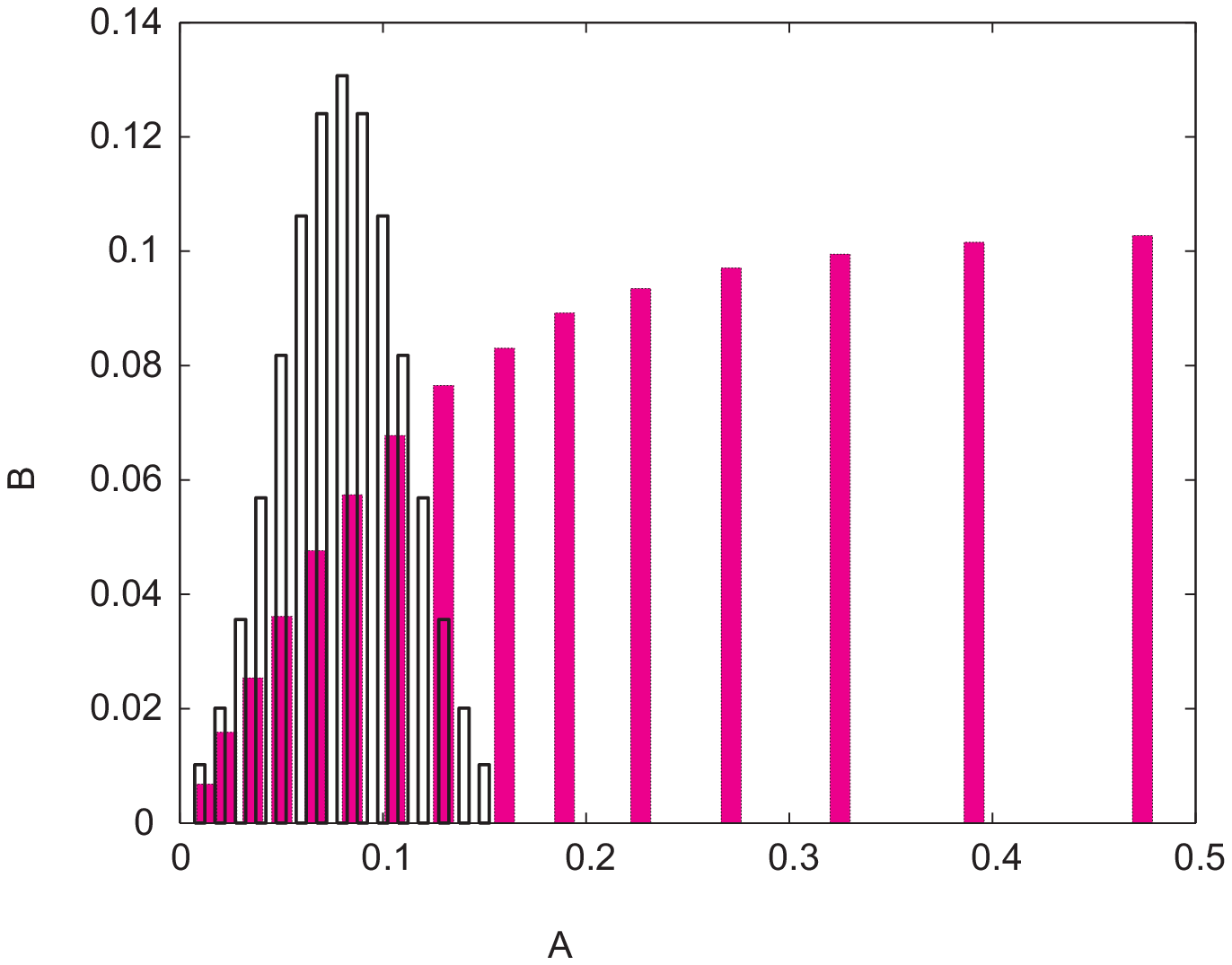}
            \end{center}
        \caption{Development of an initial \textsc{Gauss}ian void distribution after 1, 2, 5 and 10 thermal cycles.}%
        \label{Fig:BarDistribution}
        \end{figure}

We start with a discretized
\textsc{Gauss}ian distribution of 15 different void radii, $ a=10 \dots 150$~nm.
Within several stages of  thermal cycling we obtain the temporal development of
$\dist$ as illustrated in Figure \ref{Fig:BarDistribution}. 
The initially symmetric distribution changes to an
asymmetrical distribution in such a way that the fraction of
smaller voids decreases and the bigger voids grow.
Such results are well known from so-called
LSW-theories for \textsc{Ostwald} ripening, \cite{Wagner1961},
where bigger ``grains'' grow at the expense of the smaller ones due
to the \textsc{Gibbs-Thomson} effect. In our model this effect also appears
in the first stages of void growth, in which vacancy diffusion
dominates. During proceeding growth the voids reach a size, for which
the evolution is characterized primarily by elastic-plastic
deformation. For such stages the distribution function considerably
differs from a typical LSW distribution. In particular the number
of large voids extremely exceeds the number of small voids.

    \begin{figure}[htb]
            \begin{center}
            \psfrag{A}{\footnotesize \hspace{-0.0cm}\vspace{5cm} $a$ in $10^{-4}$ mm}
            \psfrag{B}{\footnotesize \hspace{-0.6cm}distribution $\dist$}
            \psfrag{C}{\footnotesize initially}%
            \psfrag{D}{\footnotesize 1 thermal cycle}%
            \psfrag{E}{\footnotesize 2}%
            \psfrag{F}{\footnotesize 3}%
            \psfrag{G}{\footnotesize 5}%
            \psfrag{H}{\footnotesize initially}%
            \psfrag{I}{\footnotesize 3}%
            \psfrag{J}{\footnotesize 6}%
            \psfrag{K}{\footnotesize 9}%
            \psfrag{L}{\footnotesize 12}%
            \includegraphics[width=0.34\textwidth]{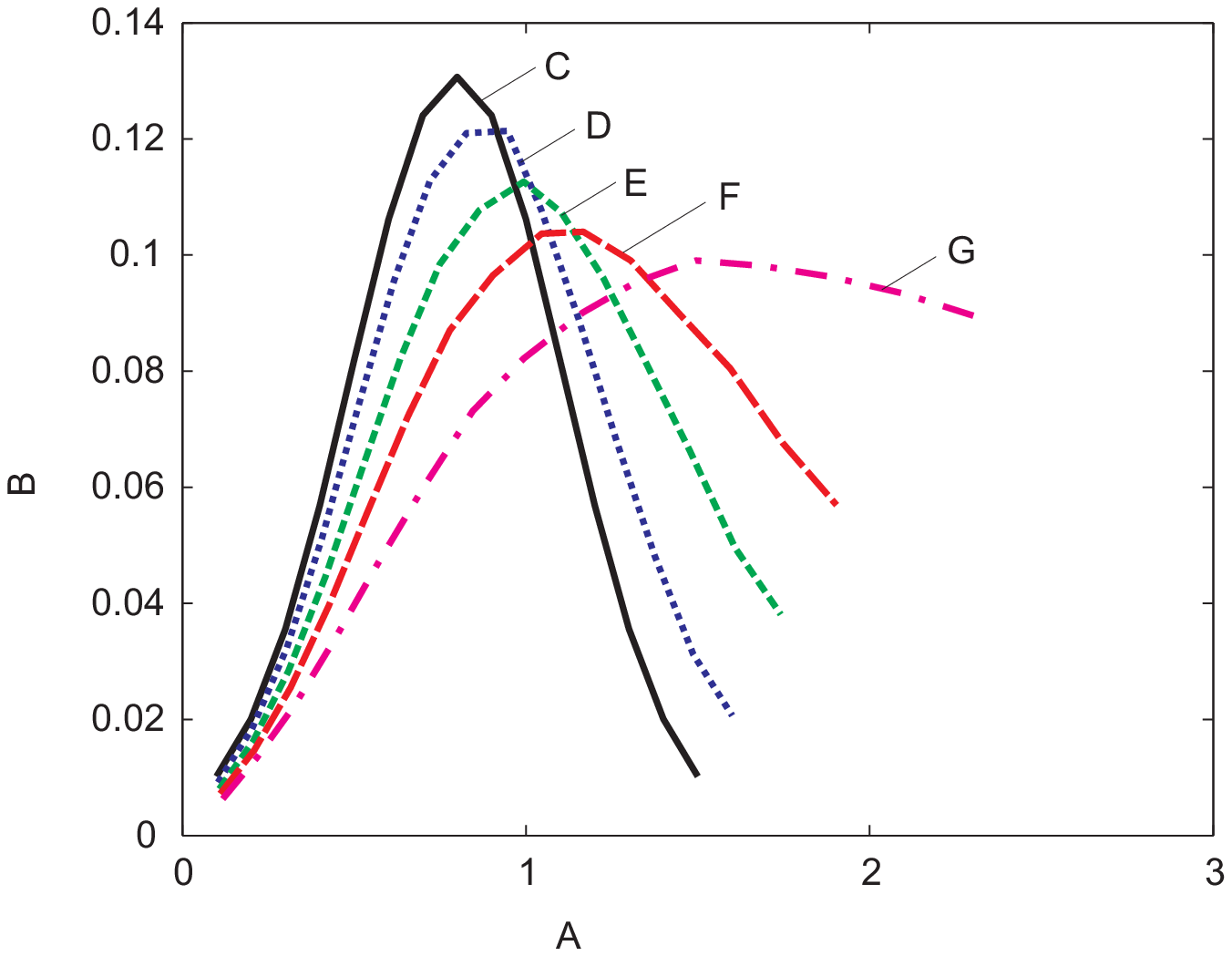}\\[5pt]
            \includegraphics[width=0.34\textwidth]{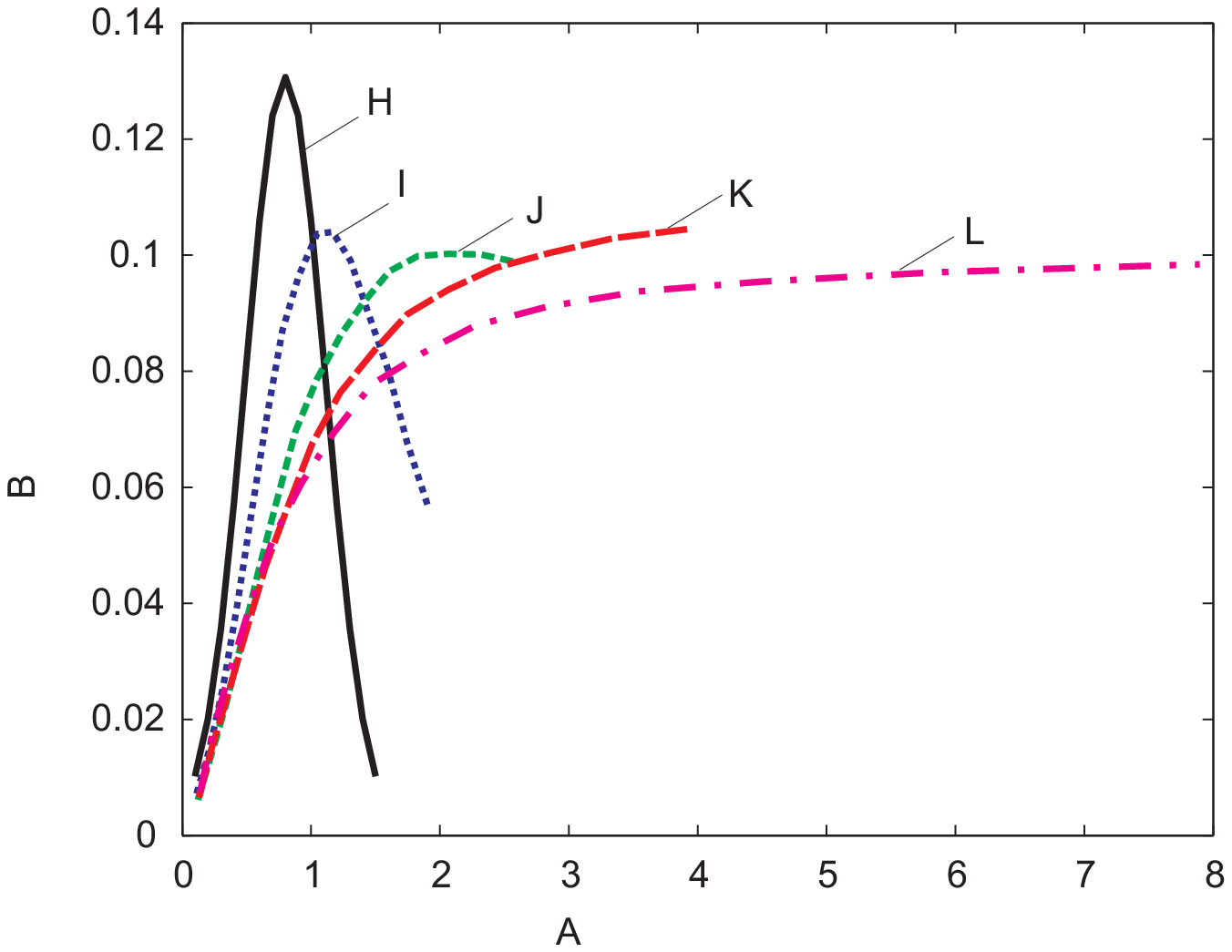}
        \end{center}
    \caption{\emph{1st row}: short time behavior of $\dist$. \emph{2nd row}: long time behavior of $\dist$.}%
    \label{Fig:ContiDistribution}
    \end{figure}

A different view on the evolution of void size distribution shows
Figure \ref{Fig:ContiDistribution}. The first
plot displays the size distribution of small voids in the
beginning of thermal loading.  
In the long range (second plot in Figure
\ref{Fig:ContiDistribution}) the growth of the bigger voids driven
by elastic-plastic deformation by far exceeds the growth of small
voids. The final distribution substantially differs from the initial
\textsc{Gauss}ian form. In particular the void regimes primarily
consists of large voids, which may allow the assumption of crack
formation by void coalescence in the immediate future (\emph{cf.},
Figure \ref{Pic:micrographs01}, \emph{2nd row}).

\begin{figure}[htb]
            \begin{center}
            \psfrag{A}{\footnotesize \hspace{-0.0cm}\vspace{5cm} $a$ in $\mu$m}
            \psfrag{B}{\footnotesize \hspace{-0.2cm}distribution $\dist$}
            \includegraphics[width=0.34\textwidth]{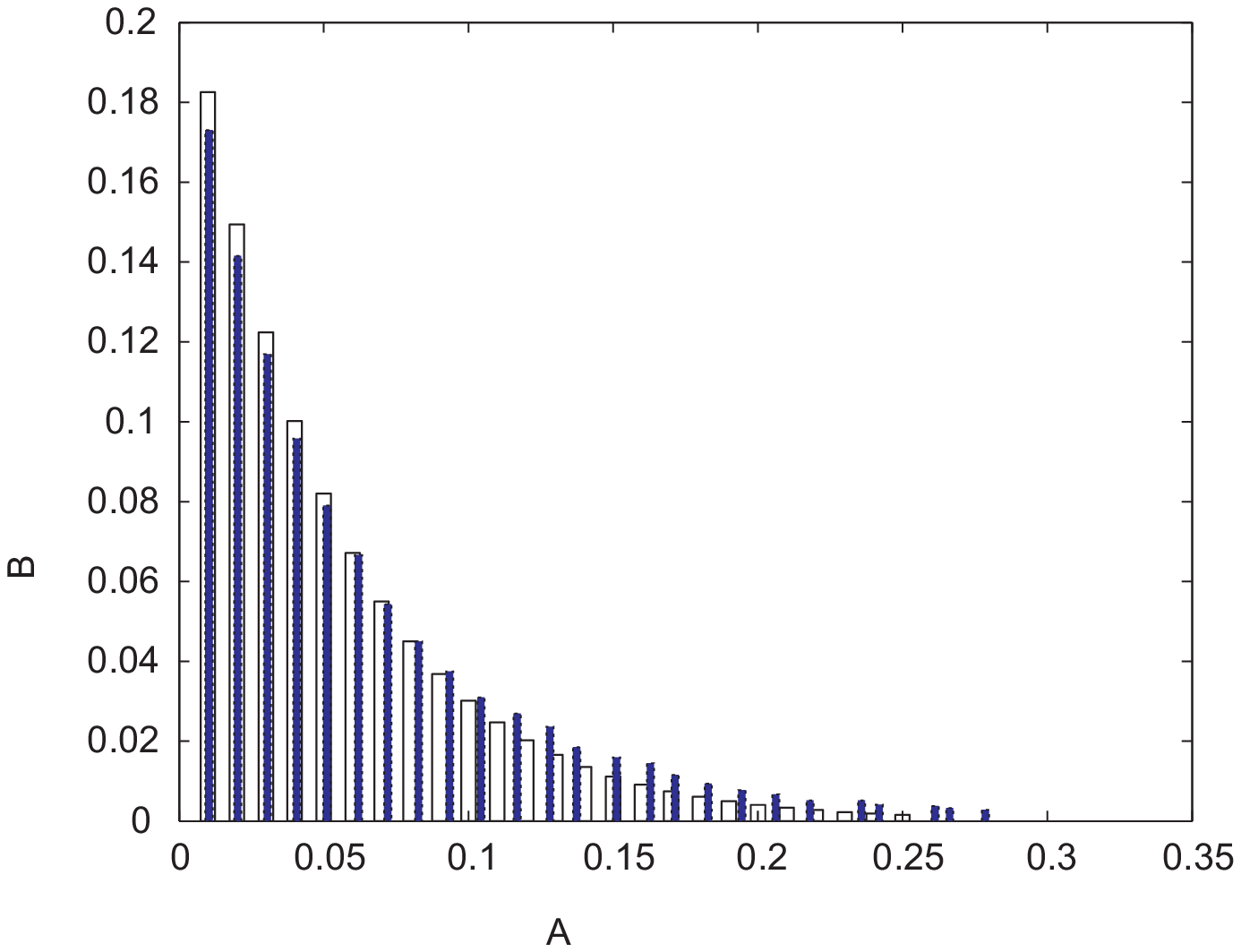}\\[5pt]
            \includegraphics[width=0.34\textwidth]{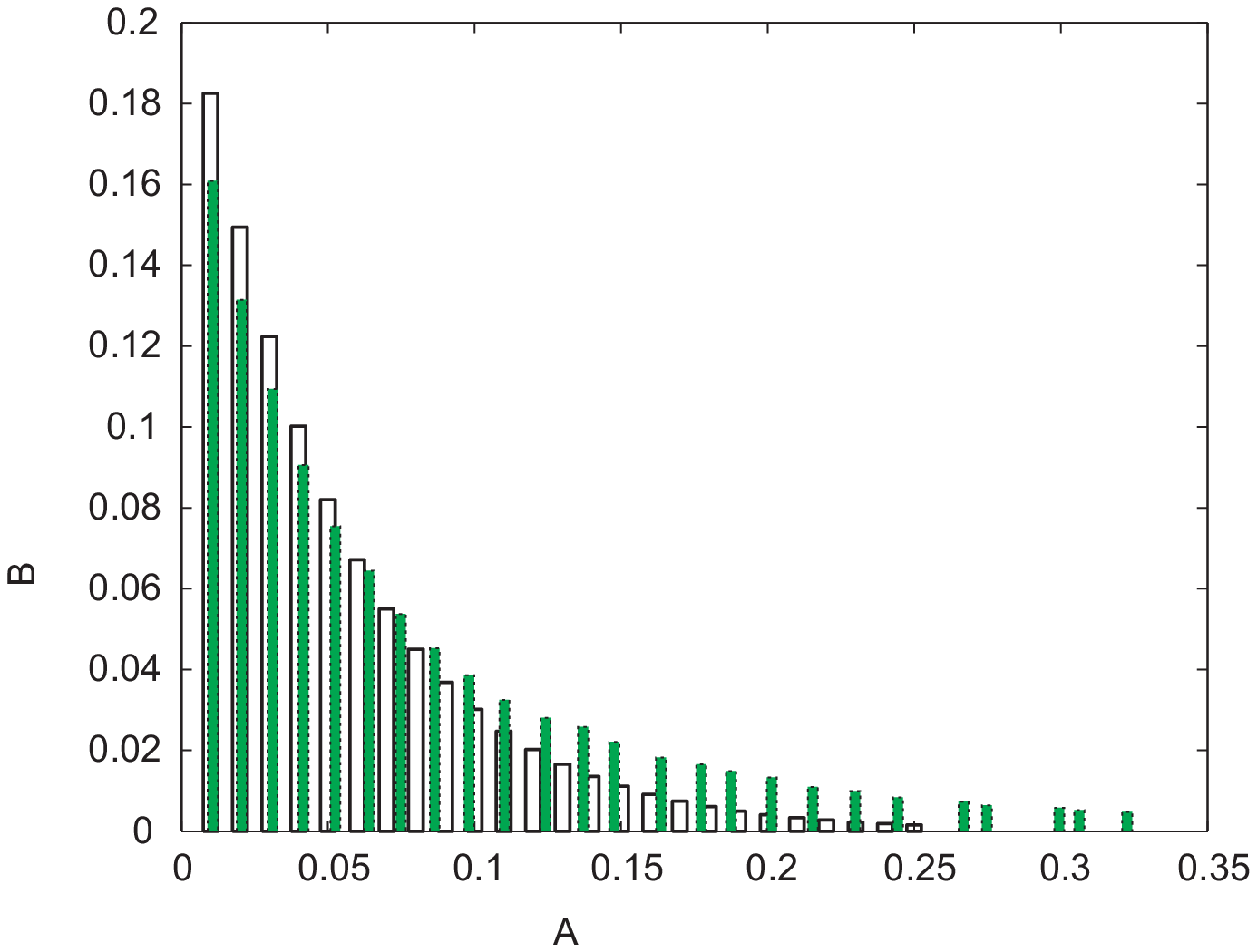}\\[5pt]
            \includegraphics[width=0.34\textwidth]{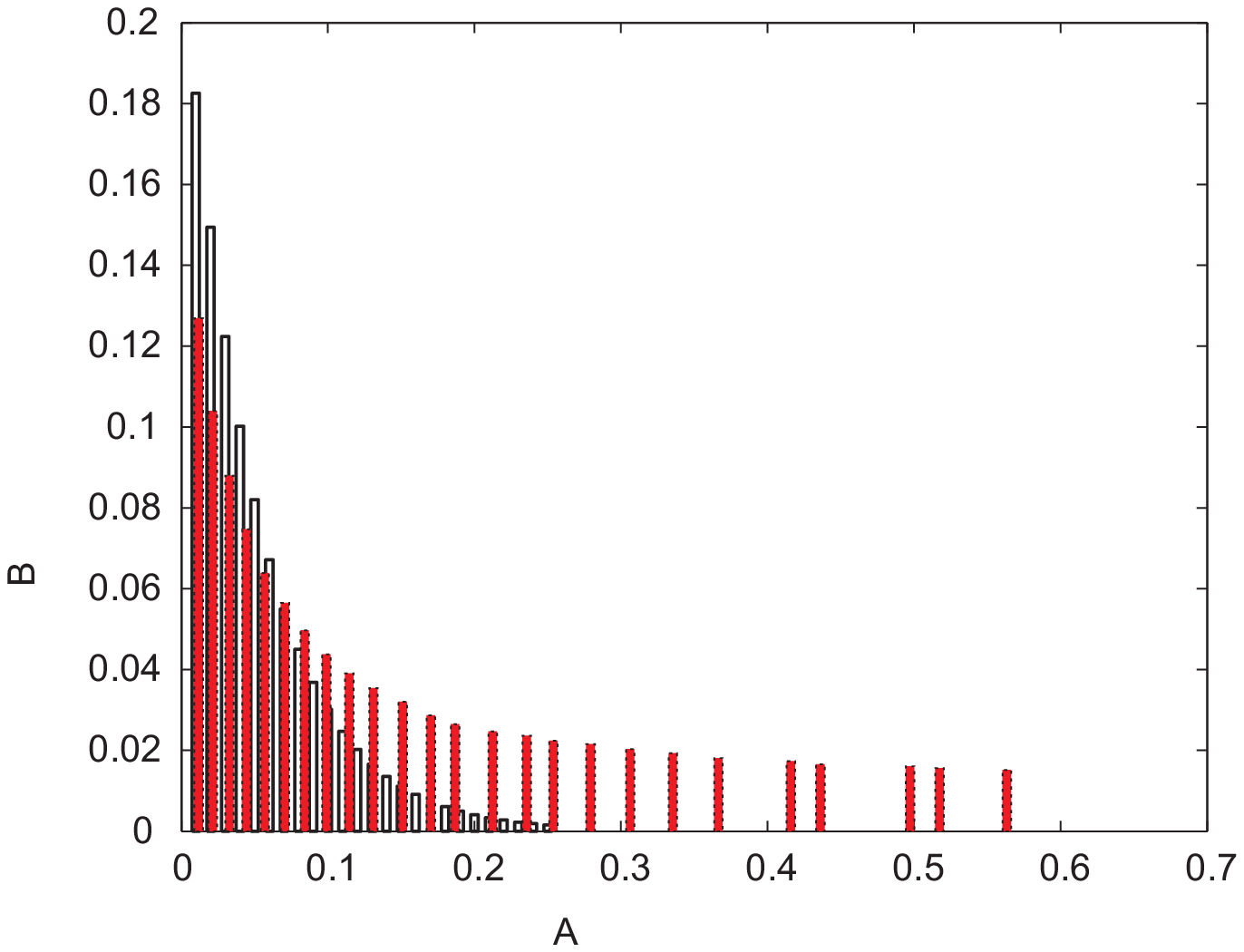}\\[5pt]
            \includegraphics[width=0.34\textwidth]{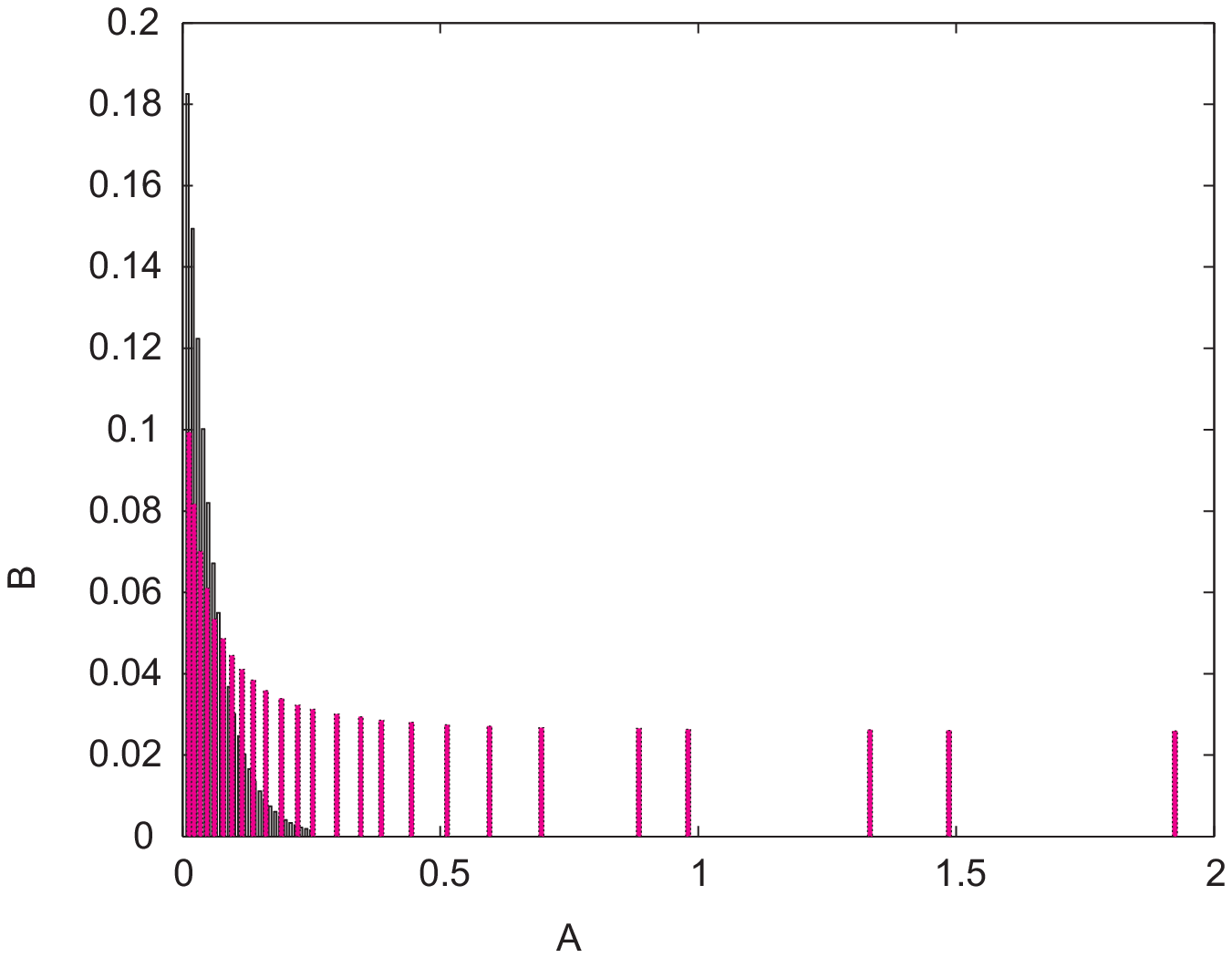}
        \end{center}
    \caption{Development of an initially \emph{exponentially decreasing} void size distribution after 1, 2, 5 and 10 thermal cycles.}%
    \label{Fig:BarDistributionExponential}
    \end{figure}

\begin{figure}[htb]
            \begin{center}
            \psfrag{A}{\footnotesize \hspace{-0.0cm}\vspace{5cm} $a$ in $\mu$m}
            \psfrag{B}{\footnotesize \hspace{-0.2cm}distribution $\dist$}
            \psfrag{6}{\footnotesize initially}%
            \psfrag{2}{\footnotesize 1 thermal cycling}%
            \psfrag{3}{\footnotesize 2}%
            \psfrag{4}{\footnotesize 5}%
            \psfrag{5}{\footnotesize 10}%
            \includegraphics[width=0.36\textwidth]{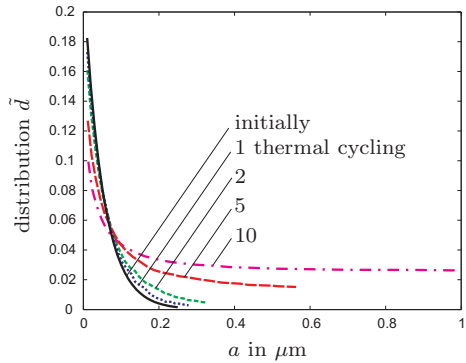}
        \end{center}
    \caption{Compact illustration of the evolution of the distribution function for different numbers of thermal cycles.}%
    \label{Fig:ContiDistrExpAll}
    \end{figure}

At second we study a distribution of void sizes, $\dist$, which --
in a more realistic manner -- accounts for the dominance of small
voids in the initial state. To this end we assume an exponential
distribution of 25 voids with $a=10 \dots 250$~nm. As displayed in
Figure~\ref{Fig:BarDistributionExponential} and
\ref{Fig:ContiDistrExpAll} we observe again a growth of the fraction
of big voids on cost of the smaller ones. In particular, the void
size $a$ reaches large values and the distribution function is
stretched over a wide range of void sizes. Here we stopped the
simulations after a  void volume fraction of $\approx\frac{2}{3}$ is
reached. Note that our model does not account for void growth due to
coalescence. Unfortunately this limits the predictive capabilities
of the void size distribution analysis. However, already a
significant amount of large voids in a material clearly indicates
the onset of failure.

\section{Conclusions}

We presented a constitutive model to predict the condensation
and growth of  \textsc{Kirkendall} voids in elastic-plastic metals,
in particular in the IMCs occurring  at the interface of microelectronic solder
joints. To this end the influence of vacancy
diffusion, surface energy, elastic-plastic and creep deformation  on the evolution of
void ensembles  during thermal cycling was investigated. %
It turns out that nano-voids collapse, whereas  voids which are
small but exceed a
critical radius (of a few vacancies) grow driven by diffusional effects. 
On the other hand the growth of
bigger voids is primarily driven by elastic-plastic deformation of the void surrounding material.
We found that work hardening plays a minor role and, as expected,
creep decelerates the void growth.

Furthermore we studied the temporal development of void ensembles
under thermal cycling. Here the presented model for void growth
is employed to (numerically) solve the balance 
of the void size distribution function. We found
that the evolving distribution initially resembles the ones known
from LSW theories, whereas the distribution during proceeded loading
evolves such that the amount of large voids drastically
increases. Such behavior correlates to experimental studies on
\textsc{Kirkendall} voids. 

The presented results are applicable to derive the temporal evolution
of the effective properties of intermetallics. The constitutive model shall be
incorporated in Finite-Element Analysis tools on the material point level to
study the mechanical behavior of IMCs under realistic loading regimes. Such analyzes
then allow for a better prediction of live time and strength 
of microelectronic joining connections.

\section*{Acknowledgment}
The authors gratefully acknowledge the support of the
German Federal Environmental Foundation (DBU).


\end{document}